\documentclass{elsart}
\usepackage[fleqn,intlimits]{amsmath}
\usepackage[dvips]{graphicx}
\begin{document}
\runauthor{G. Schmid, I. Goychuk, P. H\"anggi}
\journal{Physica A}

\begin{frontmatter}
\title{Channel noise and synchronization in excitable 
membranes}

\author{Gerhard Schmid},
\author{Igor Goychuk},
\author{Peter H\"anggi\corauthref{cor1}}
\address{Universit\"at Augsburg, Institut f\"ur Physik,  
Theoretische Physik I, 
 D-86135 Augsburg, Germany}
\corauth[cor1]{Corresponding author, tel.: +49-821-598-3250,
fax:+49-821-598-3222, e-mail: Peter.Hanggi@physik.uni-augsburg.de }

\begin{abstract}
Using a stochastic generalization of the 
Hodgkin-Huxley model, we consider the influence 
of intrinsic channel noise  
on the synchronization between the spiking activity of the excitable
membrane and an
externally applied {\it periodic} signal. 
For small patches, i.e. when the channel 
noise dominates the
 excitable dynamics, we find the phenomenon of intrinsic 
coherence resonance. In this case, the relatively regular 
spiking behavior is practically independent of the applied 
external driving; therefore no synchronization 
occurs.  Synchronization takes place, however, only for 
sufficiently large ion channel 
assemblies. The  
neuronal signal processing is thus likely rooted in the collective 
properties of optimally large assemblies of ion channels.  
\end{abstract}
\end{frontmatter}

\section{Introduction}
\label{sec:introduction}

A fundamental question in neurophysiology concerns the 
limiting factors of the reliability of 
neuronal responses to given stimuli. In this 
article we focus on a particular aspect of this 
complex issue: the impact of channel noise, which is 
generated by random gating dynamics of the 
ion channels in membrane patches of finite size. 
In particular, we investigate the effect
of channel noise on the synchronization between the 
action potential produced by the cell membrane
patch of finite size and the applied periodic stimulus.
 
The topic of synchronization, especially in biological 
systems, attracts ever growing interest,
see Refs. 
\cite{Pikovsky2001,Anishchenko2001,Freund2003,Singer1999}. 
Some prominent examples are 
the collective flashing of fireflies 
\cite{Morse1916}, the synchronization between the respiratory and
cardiac activity in human cardiorespiratory system 
\cite{Schaefer1998}, and the signal processing in sensory systems 
\cite{Neiman1999-1,Tass1998}. Recently, the phenomena of frequency
and phase synchronization have been explored in presence of ambient
noise sources, both for a noisy (overdamped) relaxation dynamics
\cite{Freund2003,Freund2000,Reguera} and for an oscillatory stochastic 
dynamics \cite{Freund2003,Callenbach2002}. 
The theoretical research is further inspired by experimental activities,
and vice versa.
In the context of theoretical modeling, 
the frequency and phase synchronization 
\cite{Rosenblum1996} has been observed, 
for example, in the integrate-and-fire model of the excitable dynamics 
driven by white noise and an externally applied stochastic 
spike train \cite{Neiman1999-2}. For an optimal 
dose of noise the mean firing rate of the driven neuron 
becomes locked by the mean frequency of the external 
spike train. 

\section{The Hodgkin-Huxley model}
\label{sec:model}

Our starting point is the well-established model of Hodgkin 
and Huxley \cite{Hodgkin1952}. The membrane patch of
area $S$ is considered as an electrical capacitor possessing 
the specific area capacitance $C$. The membrane separates two ionic 
bath solutions 
(which in vivo correspond to the interior and the exterior of the 
excitable cell)  with  different concentrations of the ions of
different sorts,  mainly potassium, ${\rm K}^+$, sodium, ${\rm Na}^+$,
and chloride, ${\rm Cl}^-$ ions. The macroscopic concentration differences 
are kept approximately constant. In the cell, this task is accomplished
by the ATP-driven ionic pumps. 
Furthermore,  the ionic baths are on the average electrically neutral. 
However, due to the different 
ionic concentrations on the opposite sides of the semi-permeable
membrane, the membrane becomes charged.  As a consequence, an
equilibrium transmembrane electrical potential difference emerges. The
lipid membrane creates an almost impenetrable barrier for the ions.
However, they can flow across the membrane  through special ion
selective pores created by specialized membrane proteins -- 
the ion channels \cite{Hille1992}. The specific potassium, $I_K$, and
sodium, $I_{Na}$, ion currents through the open ion
channels are approximately proportional to the differences of the 
transmembrane
potential $V$ and the specific (for the particular sort of ions) 
equilibrium potentials, $E_K$ and $E_{Na}$, 
respectively. The {\it stochastically} averaged, mean 
conductances, $G_{\mathrm Na}(m,h)$ 
and $G_{\mathrm K}(n)$,
are, however, strongly nonlinear functions of $V$. This nonlinearity 
emerges due to the gating
dynamics (see below). There exists also the leakage current $I_L$,
mainly due to the chloride ions. 
If the membrane is driven by the external current $I_{ext}(t)$,
the sum of the specific ion currents and the capacitive current, $I_C$,
must be equal to $I_{ext}(t)$ as a consequence of the charge conservation.
Therefore, the equation for the transmembrane potential $V(t)$ reads 
\begin{multline}
\label{eq:voltage-equation}
  C \frac{d}{dt} V +G_{\mathrm{K}}(n)\ (V-E_{\mathrm{K}}) 
  +G_{\mathrm{Na}}(m,h)\ ( V - E_{\mathrm{Na}}) \\ +G_{\mathrm{L}}
  \ (V - E_{L}) = I_{\mathrm{ext}}(t)\, .
\end{multline}
For a squid giant axon, the parameters in Eq. (\ref{eq:voltage-equation}) 
are {$E_{\mathrm Na}=50\,$mV}, 
{$E_{\mathrm K}=-77\,$mV}, {$E_L=-54.4\,$mV}, and $C=1\, \mu$F/cm$^2$.
Furthermore,  the leakage conductance is assumed to be constant, 
{$G_{\mathrm L} =0.3\,$mS/cm$^2$}. On the contrary, 
the sodium and potassium conductances
are controlled by the voltage-dependent gating dynamics of 
single ion channels and
are proportional to their respective numbers. These latter assumptions
have been fully confirmed in the
single-channel recordings by Neher, Sakmann and
colleagues which indeed have proven that ion channels undergo 
the opening-closing stochastic gating dynamics   
\cite{Sakmann1995}. In the Hodgkin-Huxley model, the opening of
the potassium ion channel is governed by four identical activation 
gates
characterized by the opening probability $n$. The channel is open
when all four gates are open. In the case of sodium channel, 
the dynamics is governed by the three independent, identical 
fast activation gates ($m$)
and an additional slow, so-termed inactivation gate ($h$). 
The independence of the gates implies that 
the probability $P_{\mathrm K, Na}$ of the occurrence of the 
conducting conformation 
is $P_{\mathrm K}=n^4$ for a potassium channel
and $P_{\mathrm Na}=m^3\, h$
for a sodium channel, respectively. In the mean-field description, 
the macroscopic potassium and sodium conductances thus read:
\begin{equation}
\label{eq:conductances-hodgkinhuxley}      
G_{\mathrm{K}}(n)=g_{\mathrm{K}}^{\mathrm{max}}\ n^{4} , \quad
G_{\mathrm{Na}}(m,h)=g_{\mathrm{Na}}^{\mathrm{max}}\ m^{3} h\, ,
\end{equation}
where {$g_{\mathrm K}^{\mathrm max}=36\,$mS/cm$^2$} and 
{$g_{\mathrm Na}^{\mathrm max}=120\,$mS/cm$^2$} denote 
the maximal conductances 
(when all channels are open). The 
two-state, open--closing dynamics of the gates 
is given by the voltage dependent opening and closing 
rates $\alpha_x(V)$ and $\beta_x(V)\; (x=m,h,n)$, i.e.  
\begin{subequations}
  \label{eq:rates}
    \begin{align}
      \alpha_{m}(V) &= \frac{ 0.1 ( V + 40 )}{1-\exp [ - ( V + 40 ) / 10] }, \\
      \beta_{m}(V) &= 4 \ \exp [ - ( V + 65 ) / 18 ]\, ,  \\
      \alpha_{h}( V ) &=  0.07 \ \exp [ - ( V + 65 ) / 20 ], \\
      \beta_{h}( V ) &= \{ 1 + \exp [ - ( V + 35 ) / 10 ] \}^{-1}\, , \\  
      \alpha_{n}( V ) &= \frac{ 0.01 \ ( V + 55 )}{ 1 - \exp [ -( V + 55 )/10 ]}, \\
      \beta_{n}( V ) &= 0.125 \ \exp [ - ( V + 65 ) / 80 ]\, . 
    \end{align}
\end{subequations}
Hence, the dynamics of the opening probabilities for the gates are given by:
\begin{align}
  \label{eq:deterministicgate}
  \dot{x} = \alpha_{x}(V)\ (1-x)-\beta_{x}(V)\ x,\quad x=m,h,n\, .
\end{align}
The voltage equation \eqref{eq:voltage-equation}, \eqref{eq:conductances-hodgkinhuxley}
and the rate equations of the gating dynamics \eqref{eq:rates}, \eqref{eq:deterministicgate}  
define the original, purely deterministic Hodgkin--Huxley model 
\cite{Hodgkin1952} for the 
squid giant axon.
 Operating with the average number of open channels, 
disregards, however, the corresponding number fluctuations 
(the so-called channel noise 
\cite{Sakmann1995,White2000}). Therefore, the Hodgkin-Huxley model is valid, 
strictly speaking,
only within the limit of very large system size where these fluctuations 
can be neglected.  
The role of internal fluctuations in 
membrane patches of finite size cannot, however, be {\it a priori} neglected.  
In fact, the recent theoretical studies make it clear that the channel noise 
can be functionally important 
for the excitable dynamics \cite{White2000,Clay1983,Schmid2001,Jung2001}.

\section{Stochastic generalization of Hodgkin-Huxley model}

\begin{figure}[t]
  \centering
  \includegraphics[width=0.8\textwidth]{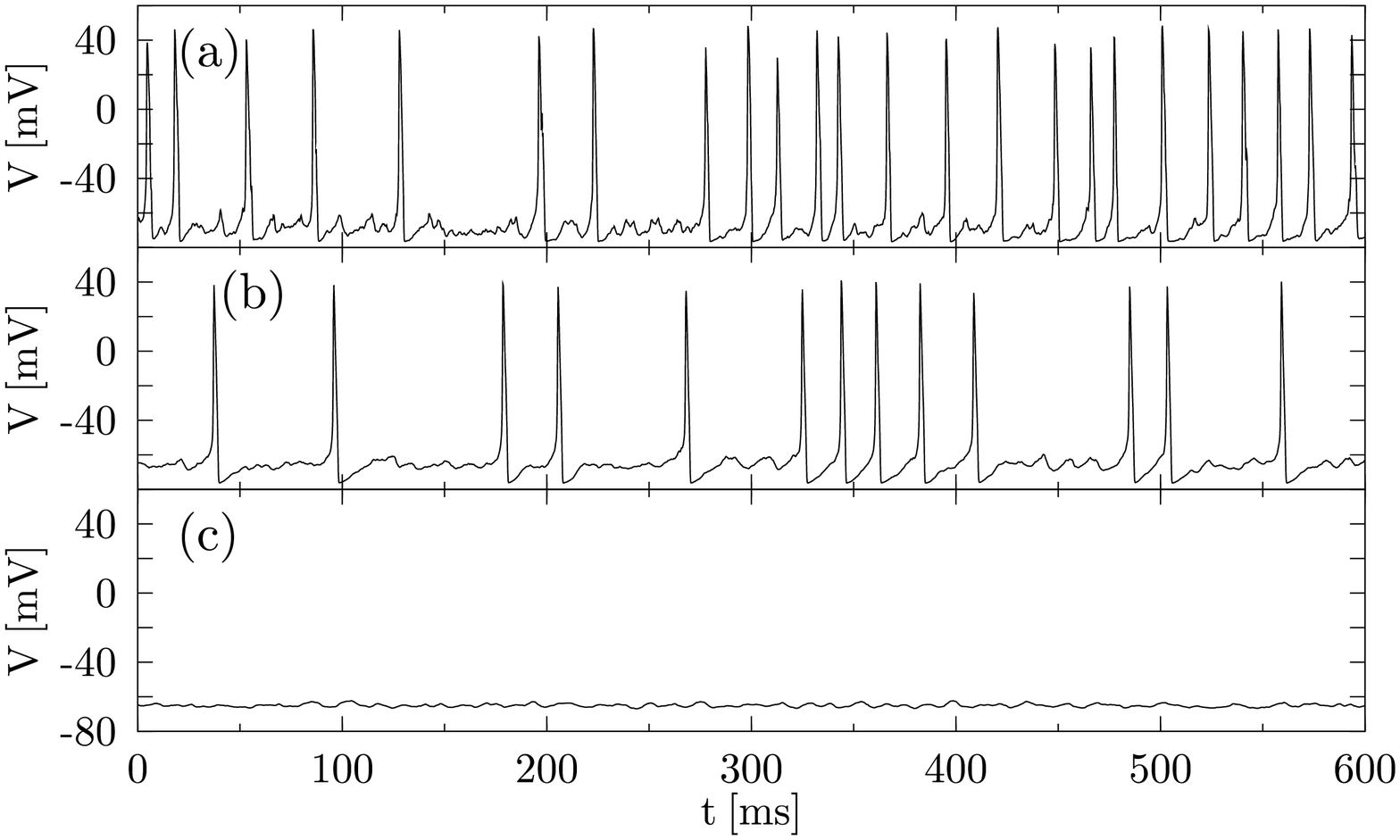}
  \caption{Three simulated realizations of voltage spike 
    trains for different patch sizes: 
    (a) {$1\; \mu$m$^2$}, (b) {$16 \;\mu$m$^2$}, and (c) {$128 \; \mu$m$^2$}. 
    With decreasing patch 
    size (bottom-to-top) the fluctuations of the number of open ion channels 
    dominate the 
    dynamics; therefore 
    more action potentials are produced, whereas for very large patch sizes  
    the intrinsic channel noise strength is not capable to produce 
    corresponding voltage spikes.}
  \label{fig:channelnoise}
\end{figure}

To account for the effect of channel noise, we use the model which presents 
a stochastic generalization of the Hodgkin-Huxley equations 
(\ref{eq:voltage-equation}-\ref{eq:deterministicgate}) due to
 Fox and Lu \cite{Fox1994}. The dynamics 
of the gating variables is  
given by the following Langevin equation:
\begin{equation}
  \label{eq:stochasticgates}
  \dot{x} = \alpha_{x}(V)\ (1-x)-\beta_{x}(V)\ x + \xi_x(t), \quad x=m,h,n\, ,
\end{equation}
with independent Gaussian white noise sources $\xi_x(t)$ of vanishing mean 
which
take into account the fluctuations of the number of open gates. 
The noise strengths
depend on the membrane voltage. The noise correlations 
assume the following form for an excitable membrane patch with 
$N_{\mathrm Na}$ sodium and $N_{\mathrm K}$ potassium ion 
channels \cite{Fox1994}:
\begin{subequations}\label{correlator}
  \begin{align}
  \langle \xi_{m}(t) \xi_{m}(t') \rangle &= \frac{2}{N_{\mathrm{Na}}}\ \frac{ \alpha_{m}
    \beta_{m}}{(\alpha_{m} +\beta_{m})}\ \delta(t -t')\, ,  \\
  \langle \xi_{h}(t) \xi_{h}(t') \rangle &=  \frac{2}{N_{\mathrm{Na}}}\ \frac{ \alpha_{h}
    \beta_{h}}{(\alpha_{h} +\beta_{h})}\ \delta(t -t')\, , \\
  \langle \xi_{n}(t) \xi_{n}(t') \rangle &=  \frac{2}{N_{\mathrm{K}}}\ \frac{ \alpha_{n}
    \beta_{n}} {(\alpha_{n} +\beta_{n})}\ \delta(t -t')  \, .     
  \end{align}
\end{subequations}
To ensure the confinement of the gating variables between 0 (all gates are closed)
and 1 (all gates are open) we implemented numerically the 
reflecting boundaries at 0 and 1.
With the assumption of homogeneous ion channel densities, 
{$\rho_{\mathrm Na} =6\cdot 10^{13}\;{\rm m}^{-2}= 60 \; \mu$m$^{-2}$}
and {$\rho_{\mathrm K} =1.8\cdot 10^{13}\;{\rm m}^{-2}= 18 \; \mu$m$^{-2}$}, the ion channel numbers are given by: 
\begin{equation}
  N_{\mathrm{Na}}= \rho_{\mathrm{Na}} S, \quad N_{\mathrm{K}}= \rho_{\mathrm{K}}
  S\, .
\end{equation}
With decreasing patch size $S$ we observe from our numerical
simulations of the Langevin equations 
(\ref{eq:stochasticgates}-\ref{correlator}) 
the more and more  increasing spiking activity due to the internal channel noise, 
see Fig. \ref{fig:channelnoise}. 
  
\section{Synchronization in excitable membranes}
\label{sec:synchronization}

\subsection{Channel noise induced spiking activity: 
Intrinsic Coherence Resonance}
\label{sec:Coherenceresonance}

\begin{figure}[t]
  \centering
  \includegraphics[width=\textwidth]{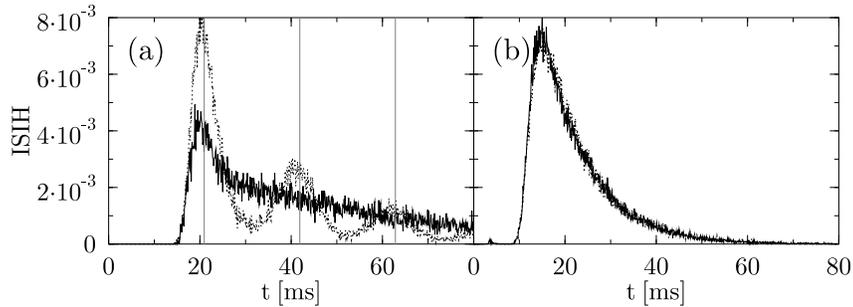}
  \caption{The interspike interval histograms (ISIH) for the undriven 
    case, $I_{\mathrm ext}=0$, (solid line) and the driven case, 
    {$I_{\mathrm ext}=\sin(0.3 \cdot t)$}, (dotted line)
    for two patch sizes: (a) {$16 \; \mu$m$^2$}, and 
    (b) {$1 \; \mu$m$^2$}. The driving independent single-peak-structure 
    becomes more and more accentuated for decreasing patch size, 
    i.e. with increasing strength of the channel noise. 
    The refractory period $T_r$ acts as lower bound to the 
    interspike intervals.  It varies with the patch area
    and becomes smaller for smaller patch sizes, cf.  $T_r\approx 15$ ms
    in part (a) vs. $T_r < 10$ ms in part (b). The driving--induced peaks appear for 
    sufficiently small internal noise at multiples of the 
    driving period $T_{driving} 
    \approx 21$ ms, cf. part (a). 
    In case of dominating channel noise, the influence of driving on the ISIH
    is barely distinguishable, cf. part (b).}
  \label{fig:interspikeinterval}
\end{figure}

We investigate numerically the influence of channel 
noise in the absence of external stimulation.
For small internal noise strengths, i.e. large patch sizes, 
the occurrence of action
potentials is very rare in the autonomous, nondriven regime: 
$I_{\mathrm{ext}}=0$, see Fig. \ref{fig:channelnoise}. 
The distribution of interspike intervals $\{T_i:=t_{i+1}-t_i \}$ is 
widely spread with longer 
intervals becoming less and less probable, see the solid line in 
Fig. \ref{fig:interspikeinterval}(a). 
With increasing channel noise the firing rate increases as well, 
and the interspike interval time--histogram depicts a more 
distinct peak structure, Fig. \ref{fig:interspikeinterval}(b). 
For small
membrane patches, the fluctuations of the number of open ion channels 
dominate the excitable dynamics and after a firing event the membrane patch 
tends to fire immediately again, but after
some refractory time which is necessary to achieve the resting 
potential after hyperpolarization. 
The above elucidation suggests that with decreasing 
patch size the spiking 
activity becomes more regular. 
A quantity that measures this regularity is the so--called
coefficient of variation ($CV$), or the relative dispersion
of the interspike interval distribution, i.e.  
\begin{equation}
  \label{eq:CV}
  CV := \frac{\sqrt{\langle T^2 \rangle - \langle T \rangle^2}}{\langle T \rangle} \ , 
\end{equation}
which involves the mean interspike interval {$\langle T \rangle := \lim_{N \to \infty} 
\sum (t_{i+1} - t_{i} )/N$} and the mean squared interval {$\langle T^2 \rangle :=  
\lim_{N \to \infty} \sum (t_{i+1} - t_{i} )^2/N$}. 
For a fully uncorrelated sequence of spikes, which corresponds to the Poissonian 
spike train, the coefficient of variation would assume the 
value $CV=1$. For a more ordered spike train, the coefficient of variation 
assumes values less then one, $CV<1$. For a purely deterministic 
signal, the $CV$ equals to zero. 

\begin{figure}[t]
  \centering
  \includegraphics[width=\textwidth]{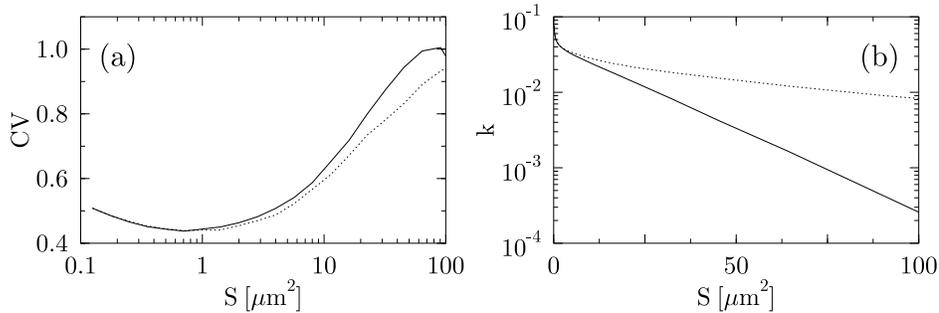}
  \caption{(a) The dependence of the coefficient of variation $CV$ in
  \eqref{eq:CV} versus patch sizes. 
    (b) Firing rates $k=1/\langle T \rangle$  
    for the undriven (solid line) 
    and driven case (dotted line, {$I_{\mathrm ext}=\sin(0.3 \cdot t)$}) 
    are depicted versus the patch areas. 
    While the firing rate is monotonically decreasing with increasing patch size the 
    coefficient of variation evidences the 
    phenomenon of intrinsic coherence resonance.}
  \label{fig:coefficientofvariation}
\end{figure}

Fig. \ref{fig:coefficientofvariation}  depicts the coefficient of variation 
\eqref{eq:CV} and the firing rate $k=1/\langle T \rangle$. While the 
firing rate decreases monotonically with respect to the
patch area, the $CV$ reveals the novel phenomenon 
of intrinsic coherence resonance \cite{Schmid2001}.
At an optimal dose of internal noise, i.e. an optimal size of the cell 
membrane patch near {$S=1\; \mu$m$^2$}, 
the $CV$ exhibits a minimum, where the spike train 
becomes distinctly 
more ordered. With decreasing 
noise level (large $S$) this feature becomes increasingly diminished,
and the $CV$ increases. 

\subsection{Sub-threshold sinusoidal driving}
\label{sec:acdriving}
Next, we switch on an external, {\it sub-threshold} sinusoidal driving, i.e. 
\begin{equation}
  \label{eq:driving}
  I_{\mathrm{ext}} := A \sin( \Omega t )\, .
\end{equation}
with $A<A_{th}$, where $A_{th}$ is the 
threshold amplitude of driving in the deterministic limit, $S\to\infty$.
For $\Omega=0.3\;{\rm ms}^{-1}$ in Fig. \ref{fig:interspikeinterval}, 
$A_{th}\approx 1.6\;\mu{\rm A/cm}^2$.
Interestingly enough the distribution of interspike intervals is 
not affected much for small patch sizes near {$S \approx 1 \; \mu$m$^2$}, 
cf. Fig. \ref{fig:interspikeinterval}(b), where the solid and dotted
curves are barely distinguishable; see also Fig. 
\ref{fig:coefficientofvariation}(a) at {$S \approx 1 \; \mu$m$^2$}. 
In this case, the spiking activity possesses an 
internal rhythm which withstands the external disturbances. For larger 
patch sizes 
the internal noise decreases and the periodic driving induces peaks at 
multiples of the driving period, see Fig. \ref{fig:interspikeinterval}(a). 
In this regime the external driving imposes a more ordered spiking activity 
which is characterized by a reduced $CV$, 
see Fig. \ref{fig:coefficientofvariation}(a). 

\subsection{The concept of stochastic phase processes: 
frequency synchronization}
\label{sec:Phaseconcepts}

\begin{figure}[t]
  \centering
  \includegraphics[width=\textwidth]{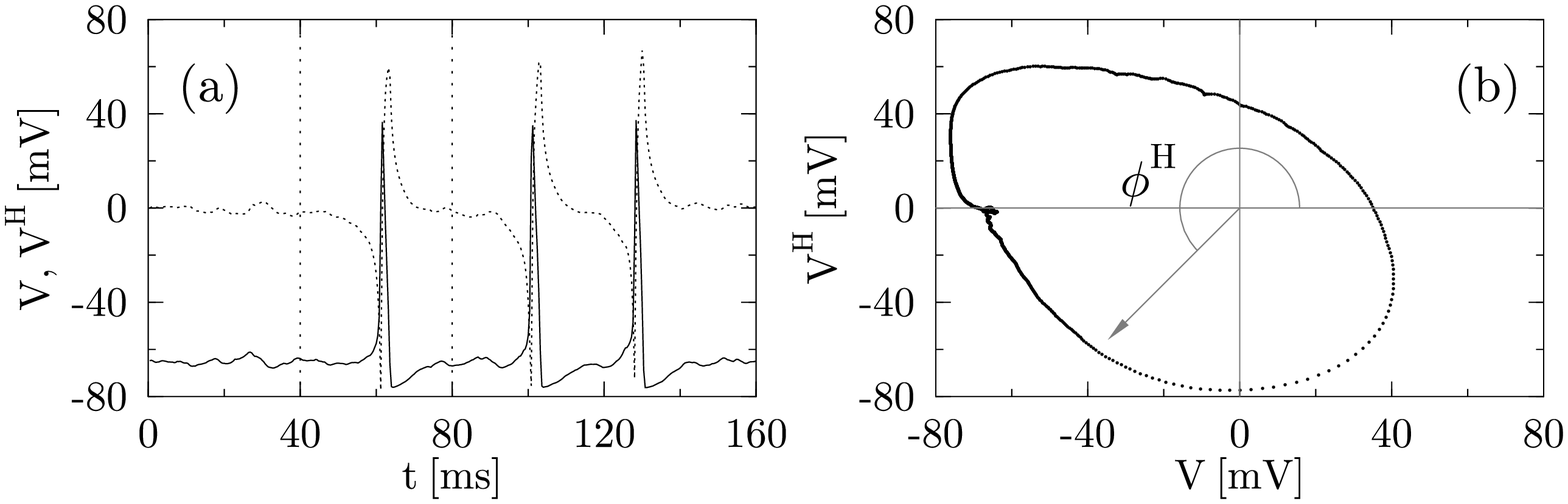}
  \caption{Stochastic (Hilbert) phase process: (a) realization of the voltage 
    signal (solid line) and the 
    corresponding Hilbert transform (dotted line); 
    (b) The trajectory in $\{V$,$V^{\mathrm H}\}$-space 
    is plotted for the time interval in 
    (a) which is determined by two dashed vertical lines at $t=40$ ms and  $t=80$ ms,
    respectively, containing a single firing event. The $\{V(t),V^{\mathrm H}(t)\}$-space 
    trajectory undergoes a full $2\pi$-revolution during such
    a single firing event.
    }
  \label{fig:hilberttransform}
\end{figure}

In order to explore the stochastic phase and and stochastic 
frequency synchronization 
between the membrane
spiking activity and the driving current, 
one has to invoke the concept of a phase process linked to
the stochastic voltage signal $V(t)$. 
The Hilbert phase $\phi^{\mathrm H}(t)$ constitutes 
one such appropriate  
phase definition. 
The corresponding approach is based on 
the Hilbert transform,  it has been originally 
introduced by Gabor \cite{Gabor1946,Panter1965}. 
The signal is extended into the complex plane with the imaginary axis 
given by the Hilbert transform of signal, $V^{\mathrm H}$, 
cf. Fig. \ref{fig:hilberttransform}:  
\begin{align}
  \label{eq:complex}
  z_{V}(t) &:= V(t) + i V^{\mathrm H}(t)\, ,\\
  \intertext{with}
  \label{eq:Hilberttransform}
  V^{\mathrm H}(t) &= H \left[ V \, \right] (t) = \frac{1}{\pi}\, P 
  \int_{-\infty}^{\infty} \, \frac{V(\tau)}{t - \tau} \, d\tau\ .
\end{align}
The integral in \eqref{eq:Hilberttransform}  is 
evaluated in the sense of the Cauchy principal value ($P$). 
The Hilbert transform is then used to define
the corresponding stochastic process for
 the Hilbert 
phase $\phi^{\mathrm H}(t)$, i.e.
\begin{equation}
  \label{eq:Hilbertphase}
  \phi^{\mathrm H} (t) = \arctan \left[ \frac{V^{H}(t)}{V(t)} \right] \ .
\end{equation}
Here, the phase should be understood as continuously growing 
function of time.
 In Fig. \ref{fig:hilberttransform}(a) 
a realization of the voltage signal and the related
Hilbert transform are plotted. Each repeated firing event 
corresponds to a cycle in the $\{V,V^{\mathrm H}\}$--space, 
see \ref{fig:hilberttransform}(b)); it thus adds an 
$2 \pi$-increment to the phase of the voltage signal. 
The corresponding Hilbert phase frequency 
$\omega_{\mathrm H}$ is then given by
\begin{equation}
  \label{eq:Hilbertfrequency}
  \omega_{\mathrm H} := \lim_{t \to \infty} \frac{\phi^{\mathrm H}(t)}{t}\ . 
\end{equation}

An alternative phase definition uses the feature of the point process that is 
generated by the spikes of the voltage signal $V(t)$. These spikes 
define marker events at times $\{t_i\}$ with each of them pinpointing 
the completion 
of a cycle. The phase increase between two subsequent marker events
is then given exactly by $2 \pi$. Moreover, by a 
linear interpolation it is possible to define the instantaneous 
phase $\phi^{\mathrm R}(t)$; i.e.,
\begin{equation}
  \label{eq:linearphase}
  \phi^{\mathrm R}(t) = 2 \pi \frac{t-t_i}{t_{i+1}-t_i} + 2 \pi i\;\;, 
  \qquad (t_i \leq t \le t_{i+1})\, ,
\end{equation}
where the times $t_{i}$ are fixed by the marker events. 
The average frequency of the 
process $\phi^{\mathrm R}(t)$ is then given by:
\begin{equation}
  \label{eq:ricefrequency}
  \omega_{\mathrm R} = \lim_{t \to \infty} \frac{\phi^{\mathrm R}(t)}{t} = 
  2 \pi \lim_{t \to \infty} \frac{N(t)}{t}\, \,, 
\end{equation}
where $N(t)$ denotes the total number of spikes within  the time interval 
$[0,t]$. The index $\mathrm R$ alludes to the so-called 
Rice frequency \cite{Freund2003,Callenbach2002,Rice1944}. 

Since each action potential adds, on the one hand, the $2 \pi$-increment 
to the Hilbert-phase and, on the other hand, defines the corresponding single 
marker event, the Hilbert frequency equals the Rice frequency, 
although the two phase definitions are different.

\subsection{Imperfect synchronization for a periodically 
driven membrane patch}
\label{sec:Frequency}

\begin{figure}[htbp]
  \centering
  \includegraphics[width=\textwidth]{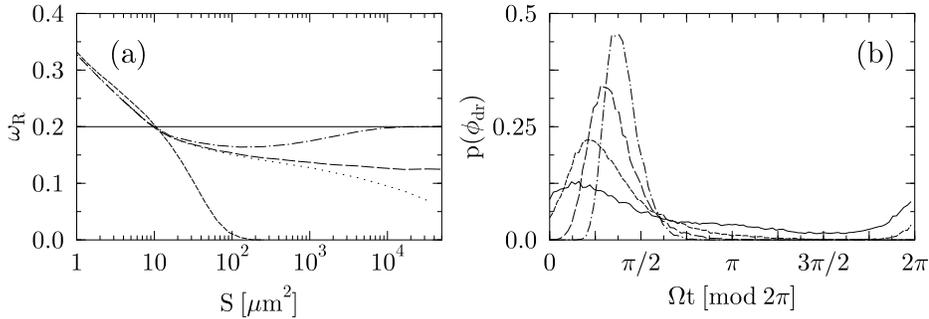}
  \caption{(a) The dependence of Rice frequency \eqref{eq:ricefrequency}
    on the patch size is plotted for three sub-threshold 
    amplitudes $A$ of the 
    sinusoidal stimulus $I_{\mathrm ext}(t) = A \sin( 0.2 \cdot t)$: 
    {$A=0.0$ $\mu$A/cm$^2$} (short--dashed line), 
    {$A=2.02$ $\mu$A/cm$^2$} (dotted line), {$A=2.05$ $\mu$A/cm$^2$} 
    (long--dashed line), and for a super-threshold amplitude 
    {$A=2.2$ $\mu$A/cm$^2$} (dotted--dashed 
    line). The threshold amplitude is $A_{th}\approx 
    2.1$ $\mu$A/cm$^2$. The level line displays the angular driving 
    frequency $\Omega=0.2\;{\rm ms}^{-1}$. 
    For subthreshold signals the firing rate vanishes for 
    large areas and the Rice frequency approaches zero, 
    while for a super-threshold amplitude the driving frequency value 
    is attained. (b) The phase probability density $p(\phi_{dr})$ of spiking
    events versus the phase $\phi_{dr}=\Omega t$ of the driving signal 
    is plotted for 
    $A=2.05 \; \mu$A/cm$^2$ 
    and four different patch areas: $S=4\; \mu$m$^2$ 
    (solid line); $S=16\; \mu$m$^2$ (short--dashed line); $S=64\; \mu$m$^2$ 
    (long--dashed line); $S=256\; \mu$m$^2$ (dotted--dashed line).}
  \label{fig:synchronization}
\end{figure}

The dependence of the mean frequency $\omega_{\mathrm R}$ on the patch 
size is depicted in
Fig. \ref{fig:synchronization}(a) for driving signals with 
frequency $\Omega=0.2$ ms$^{-1}$ and for different 
driving amplitudes $A$. In the strong noise regime, 
the spiking activity is determined 
by the internal noise and is independent of the driving  
frequency and amplitude. For intermediate-to-small noise strengths the 
spike occurrences 
are locked to {\it multiples} 
of the driving period, cf. Fig.\ref{fig:interspikeinterval}(a). 

This synchronization behavior is also reestablished 
in the phase probability density $p(\phi_{dr})$ of spiking
events in Fig.\ref{fig:synchronization}(b). 
The latter quantity is defined with respect to the phase 
$\phi_{dr}=\Omega t$ of the driving signal within the driving period,
i.e. $0\leq \phi_{dr} <2\pi$. For small patch sizes, 
i.e. for a large channel noise, the probability density $p(\phi_{dr})$ is
relatively flat. When the channel noise diminishes 
a peak structure becomes more and more pronounced. 
It worth to notice that spikes occur most frequently
{\it before} the driving current reaches its  maximum  
at $\Omega t=\pi/2$. The occurrence of the spike thus {\it forecasts}
that the signal's maximum will be reached soon. This effect is similar
to the effect of {\it anticipated synchronization} between 
master and slave
neurons \cite{calvo}. Moreover, a 
very interesting new
effect is noticeable in Fig. \ref{fig:synchronization}(b): 
the phase lag between the maximum of $p(\phi_{dr})$ and the signal 
maximum at $\phi_{dr}=\pi/2$ {\it increases} upon increasing
the channel noise strength. In other words, the smaller membrane
patch fires most probably more in advance than the larger patch, although
the probability distribution of the spikes occurrence is increasingly 
flattened. 
This prominent effect can be explained as follows. 
The increase of noise strength
certainly helps to overcome the firing threshold. Therefore, 
the system tends to fire
more and more in advance, before the maximum of signal is reached. 
However, the noise increases also the uncertainty of firing events and,
therefore, the probability distribution becomes flat. 

Even though the observed locking behavior
 presents clearly some sort of synchronization, 
no {\it perfect} frequency synchronization -- characterized 
by a perfect locking of the mean frequency \eqref{eq:Hilbertfrequency} 
to the external driving frequency -- can be detected, 
see Fig. \ref{fig:synchronization}(a). While for large, but still
sub-threshold stimuli $A<A_{th}\approx 
    2.1$ $\mu$A/cm$^2$ a plateau is formed over a wide 
range of noise strengths,
the mean frequency, which corresponds to this plateau, doesn't match the 
driving frequency. 
This frequency mismatch happens due to the multimodal 
structure of ISIH, which is caused by the locking 
of the firing occurrences to the external force period 
in ratios different from $1:1$. A similar 
phenomenon of imperfect synchronization has also been 
found in the human cardiorespiratory activity \cite{Schaefer1998}.
 The plateau 
corresponds to the situation where nearly every maximum 
of the driving force produces a spike. 
Nevertheless, some driving maxima produce no spikes and therefore
the mean frequency is smaller than the driving frequency. We point out,
however, that for a signal represented by a stochastic spike train,
like in Ref. \cite{Neiman1999-2}, a perfect synchronization should
be observed. The reason is that a random driving with the mean frequency
$\langle \Omega\rangle$ will not produce additional maxima in the ISIH
at the multiplies of the {\it mean} driving frequency. The occurrence
of such maxima thereby destroys
the perfect synchronization in the periodic case.

The situation changes with a super-threshold stimuli, see dashed-dotted
line in Fig. \ref{fig:synchronization}(a): 
here the increasing strength of the channel
noise with decreasing size $S$ destroys the perfect frequency 
synchronization. This is caused by the skipping 
of some firing events, or by the generation of
additional spikes.

\section{Conclusions}
\label{sec:Conclusion}

In conclusion, we have investigated the synchronization in 
a noisy generalization of the Hodgkin-Huxley model, which incorporates
the spontaneous fluctuations of the membrane conductivity due 
to the individual ion channel dynamics -- the so-called channel 
noise. In the absence of
 an externally applied stimulus 
the excitable membrane patch exhibits a noise-induced 
rhythmic spiking activity at an optimal patch area.
This effect can be regarded as the intrinsic coherence resonance 
\cite{Schmid2001,Jung2001}.
In the presence of external periodic driving, the 
interspike interval histograms
clearly demonstrate a synchronization of firing events with
 the external driving for large and intermediate 
patch sizes, while for small membrane patches the channel noise 
reigns the spiking dynamics. Due to the excitation of higher
harmonics, a perfect frequency synchronization could not be 
observed. Instead, the 
Rice frequency, which is defined by the occurrence frequency of  
action
potentials, exhibits a plateau-like structure for subthreshold signals: 
over a rather wide range of channel noise strength the Rice frequency 
can be fixed 
at a value smaller than the driving frequency. Furthermore, it was 
shown that the Hilbert frequency of the voltage signal equals to the
Rice frequency of the spike process and, therefore, this average Rice
frequency also reveals an
imperfect synchronization behavior.

{\bf Acknowledgments}. This work has been supported by the Deutsche
Forschungsgemeinschaft via the Sonderforschungsbereich SFB-486,
{\em Manipulation of matter on the nanoscale}, Project A10.

\end{document}